\documentclass[12pt,preprint]{aastex}

\shorttitle{The Third Galactic Quadrant}
\shortauthors{Vazquez et al.}

\begin{document}


\title{Spiral structure in the outer Galactic disk.
I. The Third Galactic Quadrant}

\author{Ruben A. V\'azquez}
\affil{Facultad de Ciencias Astron\'omicas y Geof\'{\i}sicas de la
               UNLP, IALP-CONICET, Paseo del Bosque s/n 1900, La Plata, Argentina}
\email{rvazquez@fcaglp.fcaglp.unlp.edu.ar}

\author{Jorge May}
\affil{Departamento de Astronom\'ia, Universidad de Chile,
Casilla 36-D, Santiago, Chile}
\email{jmay@das.uchile.cl}

\author{Giovanni Carraro\footnote{On leave from the Dipartimento di Astronomia, Universit\'a di Padova,
Italy}}
\affil{ESO, Alonso de Cordova 3107, Vitacura, Santiago, Chile}
\email{gcarraro@eso.org}

\author{Leonardo Bronfman}
\affil{Departamento de Astronom\'ia, Universidad de Chile,
Casilla 36-D, Santiago, Chile}
\email{leo@das.uchile.cl}

\author{Andr\'e Moitinho}
\affil{{SIM/IDL, Faculdade de Ci\^encias da Universidade de
Lisboa, Ed. C8, Campo Grande, 1749-016 Lisboa, Portugal}}
\email{{andre@sim.ul.pt}}

\author{Gustavo Baume}
\affil{Facultad de Ciencias Astron\'omicas y Geof\'{\i}sicas de la
  UNLP, IALP-CONICET, Paseo del Bosque s/n 1900, La Plata, Argentina}
\email{gbaume@fcaglp.fcaglp.unlp.edu.ar}

\begin{abstract}
  We combine optical and radio observations to trace the spiral
  structure in the Third Quadrant of the Milky Way. The optical
  observations consist of a large sample of young open clusters and
  associations, whereas the radio observations consist of a survey of
  nearby and distant clouds observed in CO. Both the optical and radio
  samples are the largest ones insofar present in the literature.  We
  use this unique material to analyze the behavior of interstellar
  extinction and to trace the detailed structure of the Third Galactic
  Quadrant (TGQ).We find that the Outer (Cygnus) grand design spiral
  arm is traced by stellar and CO components while the Perseus arm is
  traced solely by CO and is possibly being disrupted by the crossing
  of the Local (Orion) arm. The Local arm is traced by CO and young
  stars toward $l = 240^{o}$ and extends for over 8 kpc along the line
  of sight reaching the Outer arm. Finally, we characterize the
  Galactic warp and compare the geometries implied by the young
  stellar and CO components.
\end{abstract}


\section{Introduction}
The complex structure of the Milky Way in the Third Galactic
Quadrant (TGQ) was noticed time ago with the presence of an
unusual velocity field \citep{Brand1993}, and of the warp
\citep{May1988,Wouterloot1990,May1997} and corrugations that
result in the non coincidence of the \ion{H}{1} (21 cm) layer with
the formal Galactic plane ($b=0^{o}$) \citep{Burton1985}. Not long
ago, \citet{Carney1993} obtained deep optical Color-Magnitude
Diagrams (CMDs) along some lines of sight in the TGQ which allowed
to detect the Galactic warp, identified from sequences of young
stars at high negative Galactic latitudes. Recently,
\citet{Moitinho2006b} produced a map of the young stellar
component of the TGQ which provides a full 3D view of the warp.
Shortly after, \citet{Momany2006} published a detailed analysis of
the warp's structure which the reader is referred to for any
additional information.


As for the spiral structure in the TGQ, no clear spatial
description had been obtained until very recently by
\citet{Carraro2005c} and \citet{Moitinho2006b}. Using stellar
(optical) and CO data they showed for the first time the shapes of
the Outer arm and of a structure extended toward $l\sim 240^{o}$,
interpreted as being the Local arm, entering the TGQ, confirming a
previous suggestion by \citet{Moffat1979}\footnote{Strictly
speaking, the ``Outer arm'' mentioned by \citet{Moffat1979} is
closer than what is found in these studies}.  While earlier CO
data \citep{May1997} could not unveil any clear grand design arm
due to limited distance sensitivity, new deeper data
\citep{May2007} actually do show the full Outer arm in the TGQ.
This has also been very recently found in the \ion{H}{1} study of
\citet{Levine2006}, who trace spiral features up to 25 kpc from
the Galactic center. What remains unclear is the behavior of the
Perseus arm which is very well defined in the Second Quadrant
\citep[eg.][]{Xu2006} but hardly detectable in the TGQ.

From what has been delineated above, an interesting aspect emphasized
in previous investigations and which deserves a closer analysis is the
description of the Local arm in the TGQ.  On one hand, from the radio
perspective, the strong CO emission toward Vela - known as the Vela
Molecular Ridge - appears to indicate that the Local arm enters the
TGQ and extends in the direction of the Vela region
\citep[$255^{o}-275^{o}$]{May1988,Murphy1991} up to the limit of the
CO survey at 8 kpc from the Sun. Such evidence is also present in the
\ion{H}{1} study performed by \citet{Burton1985}.  On the other hand,
optical stellar data reveal the existence of an elongated structure
toward $l\sim 240^{o}$ reaching the Outer arm
\citep{Moitinho2006b}. This latter structure is also traced by CO
clouds and similarly appears to be the extension of the Local arm in
the TGQ.

In \citet{Moitinho2006b} we have presented a 3D map of the TGQ,
built from a catalog of young open clusters and fields containing
early type stars, complemented with CO observations of distant
clouds. The focus in the \citet{Moitinho2006b} Letter was on
showing how observational evidence interpreted as a galaxy being
accreted by the Milky Way - the Canis Major dwarf galaxy
\citep{Bellazzini2004,Martin2004a,MartinezDelgado2005} - was in
fact more naturally explained by the previously unclear broad
spiral structure, together with the warp, in the TGQ. In this
paper, we add new stellar fields and CO data to produce an updated
version of the map from \citet{Moitinho2006b}. In contrast with
the Letter we now focus on a detailed analysis of all the detected
spiral features and of the warp in the TGQ.  Additionally, data
analysis and the reddening pattern of the TGQ are discussed in
depth, which was not done in the previous work.

The remaining of the paper is organized as follows. In \S~2 we
present and discuss the optical sample, illustrating the details
of the derivation of fundamental parameters. \S~3 describes the
acquisition process, velocity resolution and sensitivity of the CO
molecular data sample used in this paper. The effect of reddening
and its variation across the TGQ is analyzed in \S~4. Our analysis
of the derived Galactic structure and of the warp signatures are
given in \S~5 and 6, respectively. \S~7 includes the final remarks
and the conclusions.

\section{Optical Sample}
The sample in this study is an extension of the one used in
\citet{Moitinho2006b}. With respect to that study we are adding 12
fields, where we can recognize star clusters and/or young
background populations (Blue Plumes, BPs) similar to the
ones described in \citet{Carraro2005c}, which are used for
tracing spiral features.

\subsection{The data}
This sample extension comes from an observing run conducted in
Cerro Tololo with the 1.0m telescope on the nights November 28 -
December 4, 2005. The telescope is operated by the SMARTS
consortium\footnote{http://www.astro.yale.edu/smarts/} and hosts a new
4k$\times$4k CCD camera with a pixel scale of
0$^{\prime\prime}$.289/pixel which allows to cover a field of
$20^{\prime} \times 20^{\prime} $ on the sky. All nights were
photometric with an average seeing of 0.8-1.3 arcsecs.

$UBVI$ photometry has been obtained using the
IRAF\footnote{IRAF is distributed by NOAO, which are operated by
AURA under cooperative agreement with the NSF.} packages CCDRED,
DAOPHOT and PHOTCAL following the point spread function (PSF)
method \citep{Stetson1987}. Calibration was secured through
observations of several  \citep{Landolt1992} standard fields.

Since all the nights were photometric a global photometric
solution has been derived using all the standard stars' (about
200) observations. The typical {\it r.m.s.} of the zero points are
0.04, 0.03, 0.03, and 0.03 in U, B, V and I, respectively. A
detailed description of the data reduction will be presented in
forthcoming papers (e.g. Carraro et al. (2007b)), where all the
clusters here presented will be discussed closely.

\subsection{Fundamental parameters}

Below we describe the procedures followed in deriving the
fundamental parameters -- reddening, distance and age -- of the
stellar populations found in the twelve fields listed in Table 1.
This paper is concerned with the spiral structure; therefore, from
the sample in Table~1 we only selected stellar groups younger than
100 Myr, since older groups are not useful as spiral structure
tracers. Therefore we are not using Ruprecht~1, Ruprecht~10,
Haffner11, Haffner~4, Haffner~15, Haffner~7, Berkeley~76 and
Auner~1. In some cases (Ruprecht~150 and the field centered in the
Canis Major over-density) we do not find any star cluster, but
only a diffuse young population. Photometric diagrams of all these
fields are included in a forthcoming article (Carraro et al.
2007b) but for Auner~1, which is presented in
\citet{Carraro2007a}. Notwithstanding the old clusters of our
sample are used together with the young ones to describe the
reddening pattern.

According to the age and the places where they are located,
clusters present a variety of challenges in deriving their
intrinsic parameters. In the case of very young clusters, they are
normally close to the Galactic plane where interstellar absorption
and crowded stellar fields turn membership assignments of faint
stars difficult. Yet, they contain massive (i.e. hot) stars which
can be used to derive their parameters. The classical photometric
method to address memberships in young clusters
\citep[e.g.][]{Baume1999} checks simultaneously the consistency of
the location of each star in several photometric diagrams (two
color diagrams, TCDs, and different color-magnitude diagrams,
CMDs). This method works well for stars with spectral types
earlier than $A0$ while becoming almost unpractical to segregate
late faint cluster members from field stars in crowded fields. It
has been criticized by \citet{Abt1979} but, as stated by
\citet{Turner1980} and later emphasized by \citet{Forbes1996}, it
is an efficient tool when membership estimates rely on a careful
inspection of the TCD and consistent reddening solutions are
applied. This is a crucial point, as age determinations of young
clusters may vary up to 100\% for clusters younger than $10^7$
years just because of a poor reddening determination and/or bad
membership assignments (since both affect the identification of
the turn-off point position). Even in the bright portion of the
CMDs contamination by field stars may exist. However, since the
$U-B$ index is more sensitive to stellar temperature changes than
$B-V$ \citep{Meynet1993}, the $V~ vs~ U-B$ diagram in young
clusters is quite useful for eliminating even most A-F type
foreground stars that tend to contaminate the vertical cluster
sequence in the $V/B-V$ plane. The outcome is that foreground
(colder stars in the mean) separate from hot cluster members.
Further membership control was always done if MK spectral types
were available.

Accurate distances require identifying clear sequences along a
broad range of magnitudes. To reduce field star contamination in
the lower main sequence we restrict the analysis to stars within
the cluster limits.  The limits are determined by inspection of
radial density profiles created from star counts
\citep[eg.][]{Carraro2007a} and set at the distance where cluster
stars reach the same density numbers than field stars. This
approach will normally be enough to produce clear cluster
sequences, thus allowing straightforward distance determinations.
As a final remark, the bright (massive) stars often include fast
rotators, Be-type stars and binaries (just to mention some
effects) that widen the cluster sequence and lead to poorly
constrained cluster parameters. However, the numbers of these
stars and the intensity of the phenomena are known to vary from
cluster to cluster.  Some are well known for the profusion of
peculiar stars \citep[e.g. Be stars in NGC~3766,][]{Moitinho1997},
but in most cases their numbers will be small enough so that they
will stand out from the rest of the cluster sequence and will not
affect the derived parameters as is the case of the sample
presented here.

Mean color excesses, $E_{(B-V)}$ and $E_{(U-B)}$, were
  determined by shifting the \citet{Schmidt-Kaler1982} ZAMS along the
  reddening line (given by the standard relation $E_{(U-B)} = 0.72
  \times E_{(B-V)} + 0.05 \times E_{(B-V)}^2$) until the best fit to
  blue cluster members was achieved. Errors in the mean color excess as
  listed in Table 1 were estimated by eye.

To compute cluster distances we must remove the effect of
  interstellar absorption, $A_V$. This requires the knowledge of the
  ratio of total to selective absorption, $R = A_V/E_{(B-V)}$, which
  can be higher than the normal value $R = 3.1$ in regions of recent
  star formation. Distances and ages of the youngest clusters could then be
  significantly biased by wrongly assuming the standard value. Though
  the analysis in \citet{Moitinho2001a} of the fields of several open
  clusters in the TGQ suggests that the extinction law in the TGQ is
  normal, an independent check of $R$ was performed: For every cluster
  field, the stellar distribution in the $B-V/V-I$ plane, where the
  slope of the reddening vector for normal absorbing material is given
  by $E_{(V-I)}/E_{(B-V)} = 1.244$ \citep{Dean1978}, was carefully
  examined. Since in this diagram, the slope of the stellar
  distribution and of the standard reddening line are very similar,
  deviations between them are indicative of a non standard reddening
  law. Cluster distances were derived by superposing the
  \citet{Schmidt-Kaler1982} ZAMS to the reddening and absorption
  corrected CMDs. The fitting error to each cluster was estimated by
  eye.

Cluster ages have been derived by superposing the solar
  metallicity isochrones from \citet{Girardi2000iso} onto the
  reddening-free sequences, as usual for star clusters. Emphasis was
  put in achieving the best fit for all the members located in the
  upper sequence. The age errors listed in Table 1 are more lower and
  upper limits. There are, however, cases where the fitting is
  ambiguous so that more than one isochrone was needed to trace the
  cluster sequence. This fact is assumed to be caused either by
  binarity or by the remaining material of the star formation process
  (only in case of extremely young clusters).

\section{Molecular clouds Sample in the Third Galactic quadrant}

CO surveys of the molecular cloud distribution in the TGQ of the
Galaxy have been carried out by \citet{May1997}, for d $\geq$ 2
kpc, and more recently by \citet{May2007}, for $R\geq 14.5$ kpc.
The former data set was obtained completely with the Columbia - U.
de Chile 1.2 m mm-wave telescope at Cerro Tololo; the latter one
includes observations at higher resolution and sensitivity
obtained with both the Columbia - U. de Chile telescope and the
NANTEN 4 m telescope, from the University of Nagoya, then at Cerro
Las Campanas. The Columbia - U. de Chile telescope is a 1.2 m
Cassegrain with a beam-width (FWHM) of $8.8^{\prime}$ at 115 GHz,
the frequency of the CO(J = 1-0) transition, and a main beam
efficiency of 0.82 \citep{Cohen1983,Bronfman1988}; the NANTEN is a
single-dish 4-m telescope with a beam-width of $2.6^{\prime}$ at
115 GHz, and a main-beam efficiency of 0.89 \citep{Ogawa1990}.

The survey of \citet{May1997} presented a list of 177 molecular
clouds mapped in CO, with the Columbia - U. de Chile telescope, at
a sampling of $0.25^{\circ}$ for R $\leq$ 12 kpc and of
$0.125^{\circ}$ for 12 kpc $\leq R \leq$ 14.5 kpc. The {\it
r.m.s.} noise temperature of the spectra, in $T_{a}$ units, is
better than 0.1 K at a velocity resolution of 1.3 km/s. The new
catalog of \citet{May2007} presents 46 molecular clouds with
distances beyond 14.5 kpc from the Galactic center.  These distant
clouds were mapped, with the Columbia - U. de Chile telescope, at
a sampling interval of $3.75^{\prime}$ (full sampling). The
velocity resolution of the spectra was improved to 0.26 km/s, and
spectra were integrated to obtain an {\it r.m.s.} noise
temperature per channel, in $T_{a}$ units, better than 0.18 K,
corresponding to 0.08 K when smoothed to a velocity resolution of
1.3 km/s. The overall sensitivity of the new maps is therefore,
when smoothed to $0.125^{\circ}$,  2.5 times better than for the
\citet{May1997}.

The catalog of 46 clouds mapped in CO by \citet{May2007} contains
14 sources that were either unresolved by the Columbia - U. de
Chile telescope beam or whose detection was dubious because of
poor signal to noise ratio. These 14 sources were then mapped with
the NANTEN telescope, confirming their detection and obtaining
their physical parameters. The NANTEN observations were carried on
with a sampling interval of $2.5^{\prime}$, a typical {\it r.m.s},
in $T_{a}$ units, of 0.17 K, and a velocity resolution of 0.1
km/s. Therefore the spatial resolution of the NANTEN maps is 3
times better than the Columbia - U. de Chile maps, with about the
same sensitivity per beam at a velocity resolution smoothed to
0.26 km/s. A detail of the main parameters of the CO surveys can
be found in Table 2. Since outside the solar circle there is no
distance ambiguity, heliocentric distances for all CO clouds in
Table 3 (that gives the basic CO clouds parameters) have been
determined kinematically, using the rotation curve  of
\citet{Brand1993}, with the galactic constants $\Theta_{\odot} =
220$  km  s$^{-1}$ and ${\rm R}_{\odot} = 8.5$ kpc.

\section{The Third Galactic Quadrant: The reddening pattern}

Figure~1 shows the plots of individual reddening values of all
clusters in our sample against Galactic longitude, latitude and
the distance to the Sun. We use different symbols to identify
young clusters (filled circles), BPs (open circles) and the rest
of the clusters (starred symbols). The upper left panel in Fig. 1
-- $E_{(B-V)}$ vs Galactic longitude -- shows just a few clusters
at $180^{o} < l< 210^{o}$, including the young one NGC 2129
\citep{Carraro2006a} and other old objects, with high reddening
(from 0.4 to 0.8-0.9 mag); the left lower panel -- distance from
the Sun against Galactic longitude -- shows they all are nearby
clusters at no more than $\approx 4$ kpc. It seems that the
reddening increases strongly beyond 4 kpc and for this reason,
farther clusters (if any) or other optical arm--tracers, are not
easily detected in this part of the Galaxy. An absorption window
then opens in the left upper panel at $215^{o} < l < 255^{o}$
within which we find reddening variations from 0.1 to 1 mag
approximately, reaching high values, above 1.2 mag (S305 and S309)
at 235$^{o}$, but with a mean of 0.4-0.5 mag along most lines of
sight. We remark the paucity in young cluster distances in the
left lower panel: there is a nearby group of young clusters with
distances up to 3 kpc, many of them falling well below the
Galactic plane (Fig. 1 right lower panel). Then, a second more
populated group of young clusters and BPs extending up to 12 kpc
appears. While the separation between these two groups is about 6
kpc at $l=215^{o}$, it decreases with increasing longitude down to
a minimum at $l=240^{o}$. This paucity is, somehow, mostly
occupied by old clusters showing a high dispersion around the
formal galactic plane at $b= 0^{o}$ as seen in the lower right
panel. Beyond $\approx255^{o}$, the situation changes again: few
young clusters remain in our sample, some of them show high
reddening values, up to 0.8 mag \citep[e.g. Pismis 8 and 13,
][]{Giorgi2005}, while others present reddening lower than 0.3
mag. The lower left panel shows they are all nearby objects too.
Something similar happens with old clusters such as Pismis 3 and
Saurer 2 which despite being close to the formal galactic plane
and the Sun are affected by the highest reddening found in this
small sector of the TGQ, above 1.2 mag approximately.

In the 1960s, \citet[][hereafter F68]{Fitzgerald1968} carried out
a large scale reddening distribution analysis using distances and
reddening values derived for isolated stars and open cluster stars
according to their spectral types and colors. Due to observational
limitations at that epoch, the study is mostly restricted to a 2
kpc radius around the Sun. His series of panels 46 to 59 plot the
reddening of the whole TGQ against the distance to the Sun. His
Fig. 4 sketches the  dust cloud distribution projected onto the
Galactic plane while his Figs. 5 to 8 depict the latitude
distribution of absorbing clouds at different distances from the
Sun. A comparison of F68 findings to ours is, therefore, pertinent
now.

Several features in common between the F68 reddening analysis and
ours should be emphasized: we can see distant objects (clusters
and BPs) in the region from $215^{o}$ to $255^{o}$ thanks to the
existence of a wide absorption window (the so-called FitzGerald
window) convincingly proved in F68. This window is limited at
$215^{o}$ by a dense cloud, labeled K in Fig. 4 of F68, that
covers from $185^{o}$ to $210^{o}$, and is followed by two other
far regions of medium absorption. The reddening introduced by this
region K, placed at less than 1 kpc from the Sun, is over 1.0 mag.
and spans the range of longitudes where just a few clusters are
included in our sample. However, even in this obscured sector, 9
HII regions were reported by \cite{Moffat1979} with distances from
$\approx 8$ to 2 kpc which are not included in our data sample to
keep data homogeneity. They are a continuation of young star
groups associated to the spiral--features revealed by the present
analysis (see Fig.~3 in \citet{Moffat1979} and compare it to our
Figs. 3 and 4 in advance). From $175^{o}$ to $185^{o}$, at
approximately 1.5 kpc from the Sun, F68 shows a region labeled L
with a reddening of up to 0.7 mag in complete agreement with the
reddening value we found for the cluster NGC 2129. Apart from this
object, we do not have young clusters at this location but other,
less young, clusters show comparable reddening values. At the side
$255^{o}$ to $280^{o}$ along the plane, another dense cloud,
labeled I in F68, precludes us from seeing far objects. In fact,
Figs. 5 to 8 in F68 show that the absorption starts increasing
strongly beyond 0.5 kpc from the Sun in this part of the TGQ.

From $215^{o}$ to $255^{o}$ (the FitzGerald window) the F68 maps
show that the reddening scarcely reaches values above 0.6 mag; in
the mean, reddening goes from 0.1 to 0.6 mag and only close to the
borders of the FitzGerald window can values be found exceeding 0.6
mag. This fact is entirely confirmed by our Fig. 1, left upper
panel, where with few exceptions, most of the clusters keep
reddening values below 0.8. Higher values as those at
$230^{o}-240^{o}$ in the same figure may be due to the presence of
other absorption regions behind, labeled X and T in Fig. 4 of F68,
located at 1 and 2 kpc approximately.  Leaving these exceptions
apart, the FitzGerald window extension in both longitude and
latitude, is not only confirmed by our sample of open clusters but
we are in the position to confirm that it extends up to 12 kpc
from the sun.

The upper right panel in Fig.~1 -- $E_{(B-V)}$ vs Galactic
latitude -- shows the trend of all our objects to keep below the
Galactic plane (negative latitudes). In particular, the bulk of
young clusters and BPs is located mostly below the plane (with a
peak at $b\approx-1^{o}$) extending down the plane, reaching
$b\approx -15^{o}$. Just one nearby young object has been found at
latitude $+4^{o}$. In the lower right panel of Fig.~1, the
clusters/BPs exhibit an interesting distribution. Some of them are
confined to a narrow band along $b \approx 0^{o}$ and can be found
even at 8 kpc from the Sun. Others that appear below the Galactic
plane ($b < -1^{o}$) are or nearby or are all at more than 4
kpc reaching distances as large as 12 kpc. Comparing all these
four panels we conclude, from an optical point of view, that
nearby young clusters are mostly found at $b < 5^{o}$ showing,
however, a clear trend to be located below the formal Galactic
plane ($b=0^{o}$). More distant young clusters and some BPs take
place along $b=0^{o}$ up to $\approx  8$ kpc from the Sun though
some few of them are found below the plane too. Nevertheless, the
remarkable feature longward 6--7 kpc is the presence of BPs that,
in the mean, are all between $-10^{o} < b < -2.5^{o}$
demonstrating the bending/widening of the distant Galactic plane
at a distance of $\approx 8$ kpc. The few young clusters in the
left lower panel clusters located as far as 9--12 kpc also follow
the pattern defined by BPs. The reddening latitude distribution
found with our clusters confirms too the findings of Figs. 5 to 8
in F68 in the sense that most of the reddening happens in the
solar neighborhood and that, below the plane, the reddening
decreases quickly as the distance increases.

The overall confirmation of the way reddening is affecting our
sample comes from Fig.~2 where the reddening of each cluster is
plotted against the distance to the Sun. Here, we find concluding
evidence that most of the reddening in the TGQ takes place in the
first kiloparsecs from the Sun. We interpret the pattern of Fig.~2
as due to quick and large variations of dust concentration with
longitude. In fact, extreme reddening values take place rapidly in
the first 3 kpc from the Sun. Three nearby young clusters show the
largest reddening, Pismis 8 and 13 toward Vela and NGC 2129 at the
beginning of the TGQ. There is then a paucity of objects, already
commented above, which extends up to 4 kpc. Thenceforth, the
minimum of the reddening of young objects slowly increases with
distance. Exceptions are two objects, one at 4 kpc and the other
at more than 10 kpc (Haffner 16 and the BP Haff7, respectively)
that show, however, reddening values similar to the ones found up
to 1.5 kpc from the Sun. The four old clusters at $d \geq 9.5$ kpc
that show very low reddening values are all at $|b|>7^{o}$. Very
high reddening values are also found for distances from 4 to 7
kpc. A couple of young objects shown in Fig. 2 at 5.5--6.5 kpc
(S305 and S309) and $l \approx 235^{o}$ have the highest values in
our sample comparable to the reddening shown by two old very
reddened clusters at more than 4 kpc which are located along the
Galactic plane at $245^{o} < l < 257^{o}$.

As a final comment, we see in Fig.~2 that some very distant
clusters, at more than 6 kpc, show minimum reddening values
increasing slowly and maximum reddening values which strongly
increase in the solar vicinity. This pattern has been already
revealed by F68. That is, our Fig.2 is, indeed, the prolongation
in distance and sum of effects shown in panels 46 to 59 of F68,
where panel 51 is at the beginning of the FitzGerald window and
panel 56 is at the end of it. His panel 54, centered in the middle
of the absorption window, shows the farthest objects at almost 9
kpc from the sun; we could extend our detections up to 12 kpc.
Therefore, our assumption that most of the reddening occurs in the
first kiloparsecs away the Sun but with a minimum value that keeps
constant with distance is a full confirmation of the early
findings of F68.

\section{The Third Galactic Quadrant: spiral structure}

Before entering in the details of the proposed spiral structure
scenario, a general comment is in order. It is important to be
aware of the uncertainties in the CO cloud kinematic distance
determination in the TGQ. Brand \& Blitz (1993) show that there
are important deviations from circular rotation in this quadrant.
The velocity residuals which reflect the presence of non--circular
motion, can be as high as 40 km s$^{-1}$ at around 240$^{o}$ in
longitude. Since circular rotation was assumed for all objects
when fitting the rotation curve, the large velocity residuals
found by Brand \& Blitz indicate the presence of non--circular
motions. Therefore, the kinematic distances obtained in the TGQ
using their rotation curve might be subjected to large
uncertainties, especially at those longitudes, like $240^{o}$,
where the velocity residuals are important. For example, at $l
=240^{o}$, deviations up to 20 km s$^{-1}$ from the circular
motion can produce uncertainties in the kinematic distances up to
at least 50\%.

We show in Fig. 3, central panel, the mass density map of CO
clouds from \citet{May1997, May2007} in the X-Y plane after a
smoothing process using the nearest neighbor method and a 0.5 kpc
kernel. Superposed to this density map we show the individual CO
clouds (gray filled squares), the young open clusters (black
filled circles) and BPs (white filled circles). All these stars
have ages less than 100 Myr (listed in Table~1) and are thus
supposed to lie not far from their birthplaces, mostly in spiral
arms. The new analyzed fields are shown with slightly large
symbols. The insert shows the CO mass scale of the figure in units
of $10^{5}~ M_{\odot}$. The X-Z (lower panel) and the Y-Z (left
panel) projections of these components, are also shown. In these
side panels, the density map is not shown since the smoothing
process gives unreal contours in the X-Z and Y-Z projections due
to the small Z range and the large X and Y ranges. For a better
understanding of the figure we have provided a longitude scale.

The X--Y projection demonstrates that the CO is distributed
throughout the whole TGQ showing a large mass concentration at $l
\simeq 240^{o}$ and $\approx8$ kpc from the Sun. Two other
prominent concentrations of CO complexes are evident, one at
$205^{o} < l < 215^{o}$ ($-4 < X< -1, ~4.5 <Y < 6$) and other at
$255^{o} < l < 270^{o}~ (-6 < X < -10, 0 < Y < 3)$. We remark the
extreme CO concentration nearby the Sun towards Vela ($l\simeq
260^{o}$) which is the Vela Molecular Ridge \citep{May1988}.

In the map of Fig.~4 (which is identical to Fig. 3) we have
superposed the logarithm spiral arm models of \citet{Vallee2005}
that represent the grand design features expected in the TGQ.

Just to illustrate the uncertainty involved in trying to fit
large-scale spiral features to our Galaxy we want mention that the
Russeil (2003) model for grand design features places the Outer
arm slightly closer and the Perseus arm slightly farther out.

\subsection{The Outer arm}

From both optical and CO observations, the Outer (Cygnus) arm in
the TGQ clearly looks like a grand design spiral feature extending
from $l =190^{o}$ to $l = 255^{o}$, confirming the findings in
\citet{Carraro2005c} and \citet{Moitinho2006b} as shown in Fig. 4.
Apart from not having optical tracers before $l =210^{o}$, both
distributions (CO and stars) are similar in shape, extension and
are spatially coincident. Furthermore, the data are well described
by the \citet{Vallee2005} model of the Cygnus arm, although a
slight tendency to deviate from the log-spiral model (being the CO
and optical data at a slightly smaller distance from the Sun) is
evident. This is part of the uncertainties in fitting the grand
design features as already said above. Russeil (2003) models would
fit better this arm than \citet{Vallee2005} models do but in such
a case other features in our scheme (e.g. the Perseus arm) will
not fit at all. Such a shearing of the external parts of arms is
often seen in other galaxies and is included in more sophisticated
models of the Milky Way \citep{Drimmel2001}.

The Outer arm, as traced by our sample coincides as well with the
\ion{H}{1} arm \citep{Levine2006}, and by extrapolating its
extension to the Fourth quadrant it appears to connect to the
distant \ion{H}{1} arm recently discovered by
\citet{McClureGriffiths2004}. At $l=210^{o}$ we find components of
this arm at the height of the formal Galactic plane, $b=0^{o}$,
but also others (S283 and S289) that are well below the plane. At
longitudes greater than $l=210^{o}$ the Outer arm is defined
basically by BP stars. Left panel of Fig.~3 shows the way this arm
starts to vertically develop and bend. The distance from the Sun
of this structure goes from $\simeq 7$ kpc at $l = 210^{o}$ to 12
kpc at $l = 255^{o}$, having a varying width of 2-3 kpc. Lower
panel, the X-Z projection, shows how the Outer arm is warped.
Summarizing, this arm enters the TGQ at $b = 0^{o}$ and stays more
or less coincident with the formal Galactic plane up to $l =
210^{o}$, where it starts bending, reaching its maximum height Z
about 1.3-1.5 kpc below the $b=0^{o}$ plane around $l = 250^{o}$.
The arm has a sizable vertical width of over half a kpc.

\subsection{The Local arm}

The Local (Orion) arm has been traced with some detail in the
First Galactic quadrant, but it is not very well known in the TGQ.
Historically, the situation is somewhat confusing with some
studies placing it toward $l \simeq 240^{o}$ and others at $l
\simeq 260^{o}$. As we shall see in this section, the source of
confusion is that there appear to be actually two distinct
structures in the Solar vicinity. Moreover, the adoption of either
direction as ``the'' Local arm is a result the particular samples
used in each of the previous studies. This is specially evident
when one considers that studies relied mostly either on optical or
on radio observations.

In the optical, \citet{Moffat1979} presented evidence that this
arm is a well confined structure located at $l = 244^{o}$ toward
Canis Major, whilst \citet{Vega1986} situate it at $l = 260^{o}$
toward Vela. Our optical data clearly show that a bridge exists
between the Sun and the Outer arm at $l = 245^{o}$ toward Canis
Major. In other words, we see young stars clusters and BPs at any
distance between the Sun and the Outer arm in this direction
\citep{Moitinho2006b} covering from $l = 230^{o}$ to $l =
250^{o}$, due to the particular location of the Sun. Furthermore,
there are CO cloud complexes distributed in a similar way. The
same happens in Figs.~3 and 4 which show how the distribution of
the clusters/BPs and CO clouds of our sample is compatible with
this structure being the extension of the Local arm into the TGQ.
Hereafter we shall refer to this structure as the Local arm.

We warn the reader, however, that the situation is more
complicated as far as CO clouds are concerned. The Perseus arm in
the Second Quadrant exhibits quite significant streaming motion.
With the data at disposal we cannot exclude that CO clouds in this
particular region of the TGQ ($240^{o}-245^{o}$) may be part of
the Perseus arm, and their diffused structures simply due to
streaming motions.

  The Local arm extends then up to $d \approx 8$ kpc and has
  its barycenter at $l \approx 245^{o}$.  The optical observations
  indicate that clusters in Local arm stay close to, or below, the
  $b=0^{o}$ plane in the solar neighborhood. Some few BPs follow this
  trend too and three clusters (Haffner~11, Ruprecht~35 and NGC ~2439)
  are clearly well below the Galactic plane. For $d \geq 8$ kpc from
  the Sun, the vertical extension of the Local arm (see BPs in Fig.~3)
  increases with the distance to the Sun and reaches the Outer arm at
  the point where it seems to show the maximum optical warping. Thus,
  the Local arm appears as an inter-arm feature, a bridge from the Sun up
  to the Outer arm resembling similar structures very common in
  face-on spiral galaxies (see e.g. the galaxy M 74 picture in \S5.3).
  We stress that the extension and shape of the distributions
  of the young stellar component and CO complexes, together with their
  spatial coincidence, fulfill the basic requirements for the presence
  of a spiral arm.

For $l \geq 250^o$ no stellar groups are found in our sample
except for a few nearby clusters.  This lack of distant tracers is
likely due to the visual obstruction caused by the Gum Nebula ($l
= 258^o , b = -2^o$) at 0.5 kpc from the Sun covering $\approx
36^o$ in the sky. The IRAS Vela shell is located in front of it
\citep{Woermann2001} while behind the Gum nebula extend the Vela
association and the very massive CO structures that compose the
Vela Molecular Ridge \citep{May1988}.

However, and despite not having optical tracers at large
  heliocentric distances for $l \geq 250^{o}$, the CO emission still
  continues and is present at several distances in the Vela direction
  with a noticeable concentration of massive clouds reaching 2-3 kpc
  from the Sun (the Vela Molecular Ridge) followed by a sparse series
  of other CO complexes reaching 12 kpc. In particular, we call
  attention to the massive complexes at $-8 < X < -10, ~1 < Y <
  3$. From a radio point of view, the Vela Molecular Ridge and the CO
  complexes beyond it exhibit the emission continuity and physical
  extension required to conform a spiral feature and have thus also
  been interpreted as the extension of the Local arm in the TGQ.  In
  this view, the Sun is inside the ``Local arm'' which is defined by
  CO alone in the range $ 250^{o} \leq l \leq 270^{o}$ (except for
  possibly 2-3 clusters within 2 kpc) and bends towards $l=260^{o}$.

Besides the sample biases mentioned in the beginning of this
section, the conclusions of early studies were even more affected
by their limited depth, which only allowed to explore with some
detail the first 3-4 kpc from the Sun. From our new map presented
in Figs.~3 and 4, it is seen how that is not deep enough. Indeed,
the Vela
  Molecular Ridge, which seemed to indicate the presence of a massive
  structure extending toward $l=260^{o}$, does not reach beyond 2-3
  kpc and although there are farther complexes more or less in the
  same direction, they are sparse and some are even likely part of the
  Perseus arm (see next Section). This is in contrast with the
  structure seen toward $l=240^{o}$ which is much better delineated
  besides also being defined by young stellar populations.

As a further remark, \citet{Alfaro1991}, using open clusters,
pointed out the existence of a depression at $l=240^{o}$, which
they called the Big Dent, an elliptical structure about 2-3 kpc
wide, extending down to 200 pc below the Galactic plane at a
distance of 2 kpc from the Sun.  This depression has been
confirmed in \citet{Moitinho2002d} and can also be identified in
the Z-X projection of the lower panel of Fig.~2 of
\citet{Moitinho2006b} and of Fig.~3 of this paper. According to
our scenario, this structure is part of the Local arm. It is
interesting to note that the depression together with the star
forming complex found near its deeper part \citep{Moitinho2002d}
fit well the description of a corrugated arm with star forming
complexes in its peaks and valleys \citep[e.g.
Carina-Sagittarius,][]{Alfaro1992}

\subsection{The Perseus arm}

This arm is well known in the Second Galactic quadrant
\citep{Xu2006}, but its existence has not been clearly established
in the TGQ after $l = 190^{o}$. NGC~2129 - close to the anticenter
- is a possible member of Perseus but the rest of optical data do
not show much evidence of the arm. On the stellar side, only a
clump of clusters at 6 kpc at $l \approx 245^{o}$ would be
consistent with belonging to the arm. However, they also fit
nicely the much more obvious structure defining the Local arm. It
is thus unclear if the clump belongs to the Local arm or to
Perseus.

The situation changes when one considers the CO observations.
Complexes appear at $l\approx 260^{o}$ ~($-6< X < -8, ~0< Y< 1.5$) and
$l\approx 215 ^{o}$ ~$(-1 < X < -2, 2 < Y < 3)$ which could be
defining Perseus. The complex at $l\approx 235^{o}$ ~($-3< X <
-4, ~2< Y< 2.5$) could also belong to Perseus although it is displaced
about 0.5 kpc toward the anticenter relative to the \citet{Vallee2005}
curve.

The presence of dust tracers together with the lack of a stellar
counterpart in certain regions of grand design arms is not
unexpected. An excellent example is the upper left of the M 74
galaxy shown in Fig. 5. Interestingly, the resemblance of Perseus
with the arm in M74 does not ends here. In fact, the region of
interest in M74 mimics our proposed scenario for the TGQ: A grand
design arm that is only defined by dust over a large extent (like
Perseus) is crossed by an inter-arm spiral feature that extends to
the outer galaxy (like the Local arm). Moreover, the crossing
region contains an enhanced number of young star clusters (as the
clump discussed above). Our data seem, therefore, to indicate that
the crossing of the Local arm is disrupting Perseus in the TGQ.
This could also explain the displacement of the CO complex at
$l\approx 235^{o}$.

\section{The signature of the warp}

Fig. 6 shows, from top to bottom, the CO complexes and
clusters/BPs in the $l, Z$ plane distributed according to
increasing distances from the Galactic center. This figure
provides a strong demonstration of the Galactic stellar and
molecular warp. In the upper panel, most clusters, BPs and CO
clouds are all confined to the formal plane ($b = 0^o$) with a
slight trend to increase the vertical dispersion for $l \geq
230^o$ approximately. The exception is the BP in the field of
NGC~2232. In the middle panel we can see a huge vertical
development of the Galactic disk with different components: CO
clouds and clusters compose a thin disk with a more defined trend
to lie below the formal plane for $l \geq 240^o$; BPs start
falling below the plane for $l \geq 220^o$ and a few clouds are
found well below the plane at $l \geq 250^o$. The lower panel is,
indeed, striking: the most remote objects (be they molecular
clouds, clusters or BPs) fall all abruptly below the plane at $l
\geq 210^o$. Although limited in longitude, this last panel
suggests that the maximum stellar warp occurs at $l = 240^o -
245^o$ while the maximum of the molecular warp seems to take place
at $l \geq 260^o$.

\section{Final Remarks and Conclusions}

We have provided a new detailed picture of the spiral structure of the
Galactic disk in the third quadrant by combining together
optical and CO observations.

The most striking structures in this zone of the Galaxy are the
Outer (Cygnus) and Local (Orion) arms. The observed Outer arm is
well matched by the extrapolation of the \citet{Vallee2005} spiral
curve and the Local arm extends toward $l \approx 245^o$ up to the
limit of both optical and CO observations, reaching the Outer arm
and confirming our previous results \citep{Moitinho2006b}.  The
more thorough analysis presented in this paper has revealed
previously overlooked evidence for the presence of Perseus in the
TGQ. Unlike the Outer and Local arms, which are defined by both CO
clouds and young stellar populations, Perseus appears to be
defined almost solely by CO. The lack of stellar tracers in
Perseus is similar to what is observed in a grand design arm in
M74. We hypothesize that the crossing of the Local arm is
disrupting Perseus in the TGQ. It is worth remarking that the
recent \ion{H}{1}
  map \citep[see Fig. ~2 in ][]{Levine2006} displays structures
  consistent with our description of the Outer, Local and Perseus arms.

We have addressed the historical discrepancy between optical and radio
studies regarding the direction of the Local arm. We find that
previous radio indications of the Local arm being directed toward $l
\approx 260^o$ are heavily based on the presence of the Vela Molecular
Ridge. This massive CO complex is however confined to 2-3 kpc from the
Sun. A few other CO clouds are also seen in the same general
direction, but they are sparsely distributed and some are likely part
of the Perseus arm.

As for the reddening distribution we are in the position to
generalize and confirm earlier findings by F68 in the sense that:
1.- most of the $E_{B-V}$ in the TGQ takes place in the first 2-3
kpc from the Sun; 2.- thanks to the longitude extension of a
window absorption we see very far objects; 3.- objects in this
window can be detected at more than 10 kpc from the Sun; 4.- the
$E_{B-V}$ in the Vela direction is simply huge precluding optical
observations of far objects behind it.

Finally, we have found clear signatures of the warp in the TGQ which
reaches at least 1 kpc below the $b=0^o$ plane. The maximum stellar
warp occurs at $l = 240^o - 245^o$ while the maximum of the molecular
warp seems to take place at a larger longitude, $l \geq
260^o$. Whether, or not, the maximum of the stellar warp is displaced
(ahead) with respect to that of the gaseous warp, requires an analysis
of the fourth quadrant. This study is now underway.

\acknowledgements RAV and GLB wish to thank the financial support
from the IALP (UNLP-CONICET, Argentina) and the PIP No. 5970 from
CONICET. JM and LB acknowledge support from the Chilean Center for
Astrophysics FONDAP No. 15010003. The work of GC was supported by
a contract from the Departamento de Astronom\'{\i}a, Universidad
de Chile. AM acknowledges support from FCT (Portugal) through
grant POCI/CTE-AST/57128/2004.


\clearpage

\begin{table}
\caption{{}Basic parameters of the new clusters and back-ground
populations} \fontsize{7} {11pt}\selectfont
\begin{tabular}{llccccccccc}

\hline\hline   \noalign{\smallskip}\multicolumn{1}{c}{Field}&
\multicolumn{1}{c}{Name} & \multicolumn{1}{c}{l} &
\multicolumn{1}{c}{b} & \multicolumn{1}{c}{(m-M)} &
\multicolumn{1}{c}{E(B-V)}  & \multicolumn{1}{c}{d$_{\odot}$} &
\multicolumn{1}{c}{Age}& \multicolumn{1}{c}{X}&
\multicolumn{1}{c}{Y}& \multicolumn{1}{c}{Z} \\
\noalign{\smallskip} \hline\noalign{\smallskip}
 && [deg] & [deg]  & mag &  mag & [kpc] & [Myr] & [kpc] & [kpc] & [kpc] \\  \noalign{\smallskip}
\hline  \noalign{\smallskip}
1&Ruprecht 1         & 223.990 & -9.690  & 12.00$\pm0.30$ & 0.30$\pm0.10$ & 1.60$\pm0.20$ & 300$\pm50$ & -0.958  & 0.993 & -0.236\\
&BP Ruprecht 1        & 223.990 & -9.690  & 15.85$\pm0.35$ & 0.40$\pm0.15$& 8.40$\pm0.70$ & $\leq100$ & -5.751  & 5.957 & -1.414\\
2&vdB-Hagen 92$^{*}$ & 224.570 & -2.490  & 10.40$\pm0.15$ & 0.15$\pm0.05$ & 0.97$\pm0.07$ &  40$\pm10$ & -0.680  & 0.690 & -0.042\\
3&Berkeley 76        & 225.099 & -1.998  & 16.50$\pm0.40$ & 0.40$\pm0.15$ &11.30$\pm1.50$ &2000$\pm250$ & -7.645  & 7.619 & -0.377\\
&BP Berkeley 76      & 225.099 & -1.998  & 15.80$\pm0.50$ & 0.50$\pm0.02$ & 7.30$\pm1.60$ & $\leq100$ & -5.168  & 5.150 & -0.255\\
4&Haffner 4          & 227.900 & -3.586  & 14.75$\pm0.20$ & 0.53$\pm0.07$ & 4.20$\pm0.50$ & 300$\pm50$ & -2.740  & 2.476 & -0.231\\
&BP Haffner 4        & 227.900 & -3.586  & 15.85$\pm0.60$ & 0.50$\pm0.07$ & 7.60$\pm1.20 $& $\leq100$ & -5.628  & 5.085 & -0.475\\
5&Auner 1            & 232.110 & -6.200  & 15.75$\pm0.15$ & 0.32$\pm0.05$ & 8.90$\pm0.60$ &3250$\pm250$ & -6.800  & 5.300 & -0.990\\
&BP Auner 1          & 232.110 & -6.200  & 16.00$\pm0.30$ & 0.40$\pm0.10$ & 8.95$\pm0.40$ & $\leq100$ & -7.022  & 5.464 & -0.967\\
6&Ruprecht 10        & 232.553 & -5.584  & 12.80$\pm0.40$ & 0.25$\pm0.15$ & 2.50$\pm0.40$ &800$\pm150$ & -1.738  & 1.331 & -0.214\\
&BP Ruprecht 10      & 232.553 & -5.584  & 15.95$\pm0.50$ & 0.50$\pm0.10$ & 7.90$\pm1.80$ & $\leq100$ & -6.242  & 4.781 & -0.769\\
7&BP Ruprecht 150    & 240.010 & -9.647  & 15.70$\pm0.50$ & 0.35$\pm0.05$ & 8.60$\pm1.80$ & $\leq100$ & -7.343  & 4.238 & -1.441\\
8&Haffner 11         & 242.395 & -3.544  & 15.25$\pm0.25$ & 0.50$\pm0.10$ & 5.50$\pm0.50$ & 800$\pm100$ & -5.307  & 2.775 & -0.371\\
9&Haffner 7          & 242.673 & -6.804  & 14.55$\pm0.25$ & 0.27$\pm0.15$ & 5.50$\pm0.50$ & 600$\pm100$ & -3.705  & 1.915 & -0.498\\
&BP Haffner 7        & 242.673 & -6.804  & 15.75$\pm0.85$ & 0.20$\pm0.10$ &10.60$\pm3.00$ & $\leq100$ & -9.351  & 4.832 & -1.256\\
10&BPCMa             & 244.000 & -8.000  & 16.04$\pm0.10$ & 0.40$\pm0.10$ & 9.30$\pm0.40$ & $\leq100$ & -8.100  & 3.900 & -1.300\\
11&Ruprecht 30$^{*}$ & 246.419 & -4.046  & 12.10$\pm0.20$ & 0.47$\pm0.03$ & 1.30$\pm0.20$ & 60$\pm15$ & -1.188  & 0.519 & -0.092\\
&SCRuprecht 30       & 246.289 & -4.030  & 17.25$\pm0.10$ & 0.55$\pm0.05$ &12.60$\pm0.40$ &  20$\pm5$ & -11.508 & 5.054 & -0.886\\
&BP Ruprecht 30      & 246.419 & -4.046  & 16.65$\pm0.15$ & 0.55$\pm0.05$ & 9.85$\pm1.30$ & $\leq100$ & -9.005  & 3.931 & -0.695\\
12&Haffner 15        & 247.952 & -4.158  & 14.20$\pm0.20$ & 0.70$\pm0.15$ & 2.50$\pm0.30$ & 600$\pm150$ & -2.219  & 0.899 & -0.174\\

\hline\noalign{\smallskip}
\end{tabular}
\tablenotetext{*} {Very preliminary values for distance, reddening
and age. Object seems to be just a sparse group of young and
mid-age stars}
\end{table}

\clearpage

\begin{deluxetable}{lcccc}
\tablewidth{0pt} \tablecaption{Main parameters of CO surveys}
\tablehead{ \multicolumn{1}{c}{} & \multicolumn{2}{c}{May et al.
(1997)} & \multicolumn{2}{c}{May et al. (2007)} } \startdata
Telescope & Col-UCH\tablenotemark{a} & Col-UCH\tablenotemark{a} &
Col-UCH\tablenotemark{a} & Nanten\\
Diameter & 1.2 m & 1.2 m & 1.2 m & 4 m \\
Telescope HPBW\tablenotemark{b} & $8.8\arcmin$ & $8.8\arcmin$ &
$8.8\arcmin$
& $2.6\arcmin$\\
Sampling & $15\arcmin$ & $7.5\arcmin$ & $3.75\arcmin$ &
$2.5\arcmin$\\
Velocity resolution (km s$^{-1}$) & 1.3 & 1.3 & 0.26 & 0.1 \\
RMS sensitivity $(\sigma[T_{A}^{\ast}])$\tablenotemark{c} & $\leq
0.10$ K & $\leq 0.10$ K & $\leq 0.08$ K & $\leq 0.07$ K\\
Number of clouds & 85 & 92 & 32 & 14 \\
Distance (kpc) & $d>2$; $R\leq 12$ & $R>12$ & $R>14.5$ & $R>14.5$\\
\enddata
\tablenotetext{a} {Columbia - U. Chile} \tablenotetext{b}{At 115
GHz} \tablenotetext{c}{At velocity resolution of 1.3 km s$^{-1}$}
\end{deluxetable}

\clearpage

\begin{table}
\caption{{}Basic parameters of the new distant CO clouds}
\fontsize{12} {10pt}\selectfont
\begin{tabular}{crrrr|crrrr}
\hline\hline   \noalign{\smallskip}

\multicolumn{1}{c}{l} &\multicolumn{1}{c}{b} &
\multicolumn{1}{c}{$V_{lsr}$} &\multicolumn{1}{c}{d$_{\odot}$} &
\multicolumn{1}{c}{Z} & \multicolumn{1}{c}{l} &
\multicolumn{1}{c}{b} & \multicolumn{1}{c}{$V_{lsr}$} &
\multicolumn{1}{c}{d$_{\odot}$} &
\multicolumn{1}{c}{Z} \\
  \noalign{\smallskip}
\hline
  \noalign{\smallskip}
  [deg] & [deg] & [km $s^{-1}$] & [kpc] & [pc] &   [deg] & [deg] & [km $s^{-1}$] & [kpc] &
  [pc]\\
  \noalign{\smallskip}
\hline
  \noalign{\smallskip}
 195.50 &  0.50& 31.73 &  10.738&    93.7&  249.41 & -2.22& 103.71 & 12.716&  -494.8\\
 195.68 & -0.12& 32.44 &  10.990&   -23.9&  249.61 & -2.10& 106.08 & 13.195&  -484.6\\
 195.93 & -0.68& 32.78 &  10.875&  -130.4&  250.00 & -3.37&  98.42 & 11.696&  -689.7 \\
 204.68 &  0.00& 45.97 &   9.407&    0.0 &  250.04 & -3.37&  97.34 & 11.502&  -678.3  \\
 208.37 & -1.81& 52.48 &   9.604&  -303.8&  250.12 & -3.12&  94.85 & 11.065&   604.0\\
 208.81 & -1.87& 50.53 &   8.710&  -285.1&  250.68 & -3.56&  89.79 & 10.238&  -637.3  \\
 211.06 &  1.18& 54.72 &   8.990&   186.3&  250.93 & -5.12&  88.80 & 10.088&  -904.7 \\
 215.87 & -2.00& 57.72 &   8.066&  -281.6&  252.04 & -4.16&  94.79 & 11.085&  -807.5 \\
 217.62 & -2.62& 63.86 &   9.068&  -415.7&  252.08 & -4.04&  88.89 & 10.138&  -716.3 \\
 218.62 &  0.25& 59.53 &   7.763&    33.8&  252.12 & -4.25& 102.49 & 12.457&  -925.7 \\
 218.75 & -0.43& 56.56 &   7.110&   -54.2&  254.62 & -0.93&  87.49 & 10.039&  -164.2 \\
 221.75 &  0.56& 63.36 &   7.896&    77.5&  254.87 & -0.93&  86.84 &  9.957&  -162.9 \\
 221.75 &  0.31& 63.25 &   7.874&    42.9&  256.18 & -1.50&  88.00 & 10.207&  -267.2 \\
 223.75 & -2.00& 63.04 &   7.473&  -260.9&  257.56 & -2.18&  94.70 & 11.323&  -432.5 \\
 226.43 & -0.18& 66.30 &   7.674&   -25.1&  257.56 & -2.18&  90.80 & 10.718&  -409.3 \\
 229.75 &  0.06& 70.44 &   7.992&     8.7&  259.16 & -1.33&  92.89 & 11.159&  -259.7 \\
 235.37 & -1.75& 81.12 &   9.304&  -284.2&  263.60 & -3.97& 111.01 & 14.674& -1020.7 \\
 235.68 & -1.18& 75.06 &   8.229&  -170.5&  263.80 & -4.14& 114.11 & 15.314& -1108.4 \\
 237.25 & -1.25& 77.34 &   8.500&  -185.4&  263.85 & -4.14& 111.26 & 14.747& -1067.4 \\
 237.25 & -1.20& 75.21 &   8.156&  -177.9&  267.41 & -4.00& 108.17 & 14.566& -1018.4 \\
 239.12 & -1.87& 83.13 &   9.359&  -306.3&  269.60 & -3.78& 108.95 & 15.010&  -992.4 \\
 244.43 & -1.93& 92.73 &  10.783&  -364.7&  271.05 & -5.51& 111.44 & 15.700& -1507.5\\
 249.18 & -3.81&  89.67 & 10.193&  -679.2&  271.45 & -2.63&  93.83 & 12.762&  -585.5\\

   \noalign{\smallskip}
\hline
\end{tabular}
\end{table}

\clearpage
\begin{figure}
\includegraphics[scale=0.8]{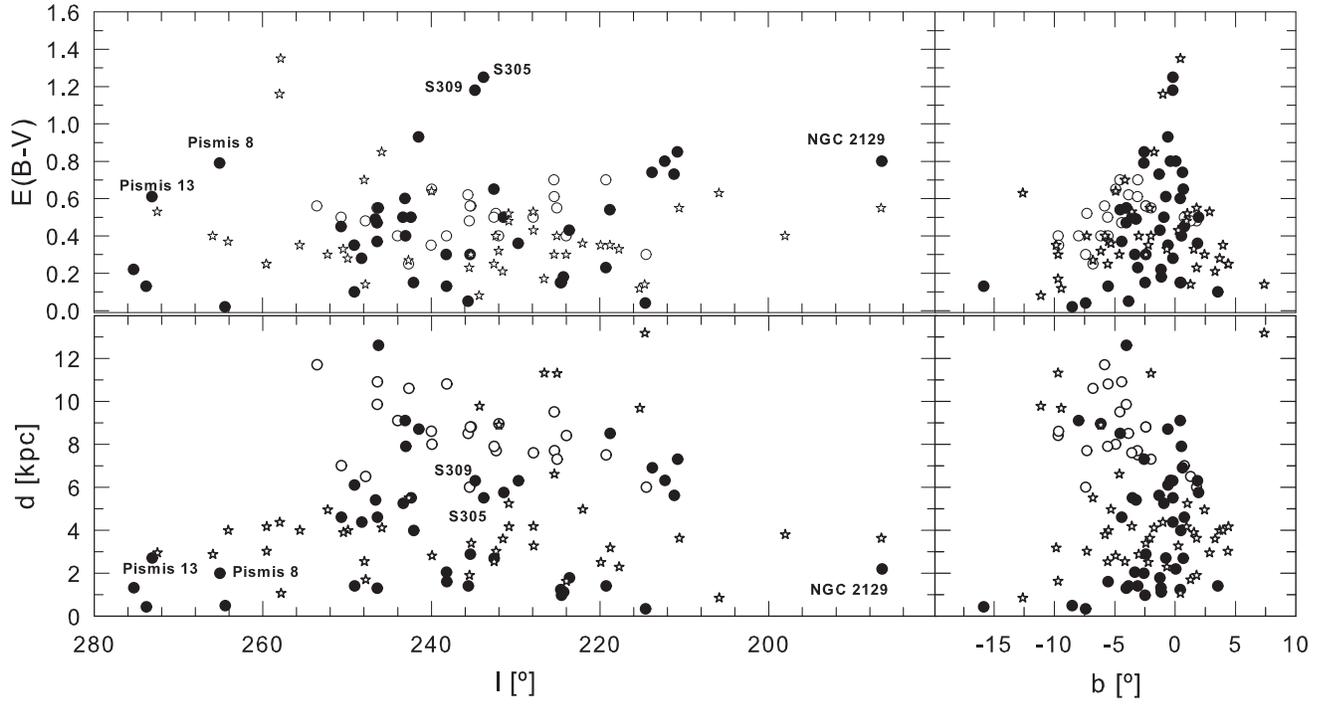}
\caption{Reddening path and distances in the TGQ: Left and right
upper panels represent the $E_{(B-V)}$ vs Galactic longitude and
latitude respectively. Left and right lower panels show the same
but for distances from the Sun. Filled and open circles indicate
clusters and BPs. Open stars show old clusters. Some objects
discussed in the text are indicated}
\end{figure}

\clearpage

\begin{figure}
\includegraphics[scale=1.4]{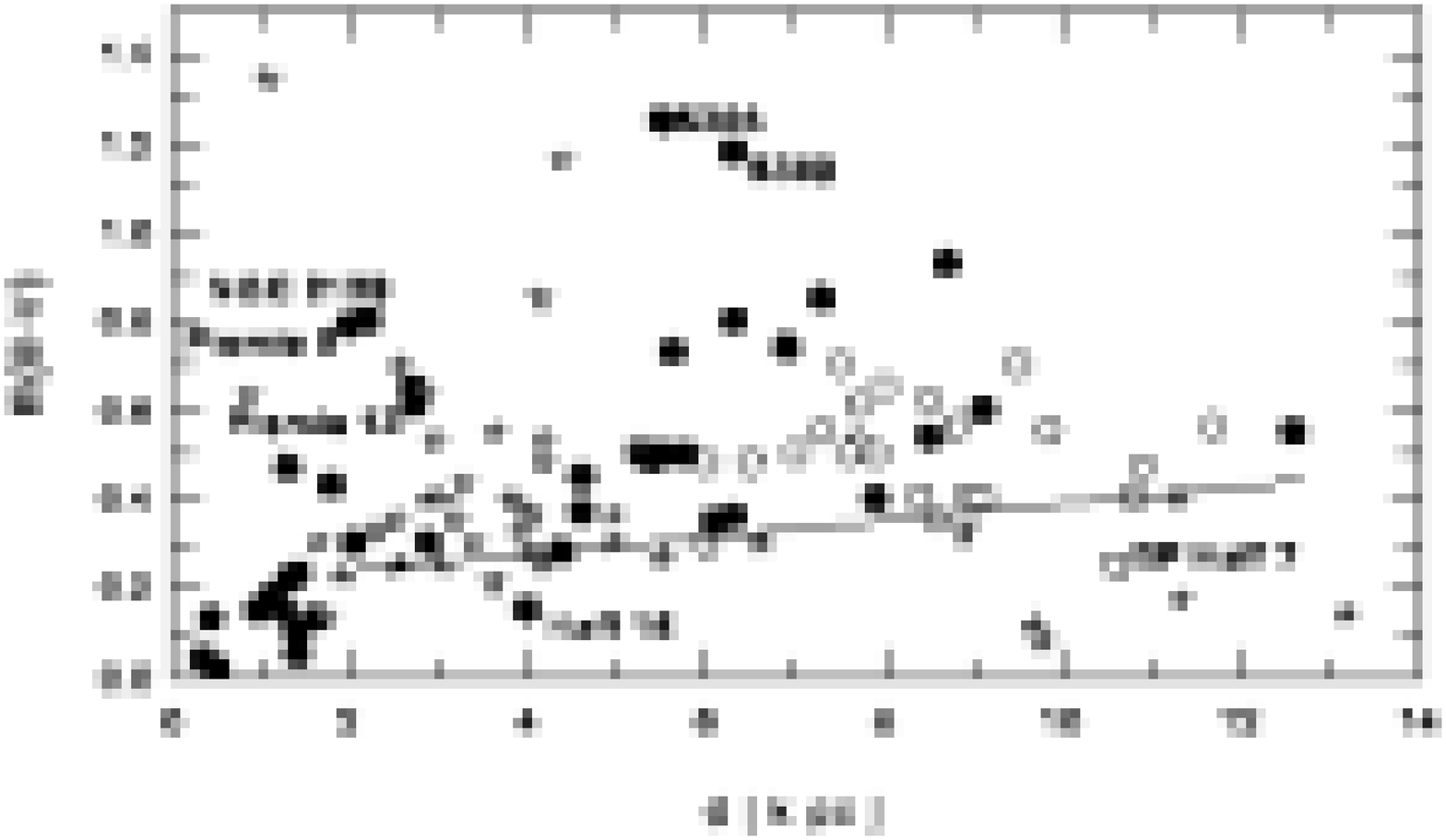}
\caption{Young and old open clusters and BPs in the
distance-$E_{(B-V)}$ plane. Symbols as in Fig. 1. Solid line
represents approximately the way the minimum reddening increases
with distance}
\end{figure}

\clearpage
\begin{figure}
\includegraphics[scale=2.2]{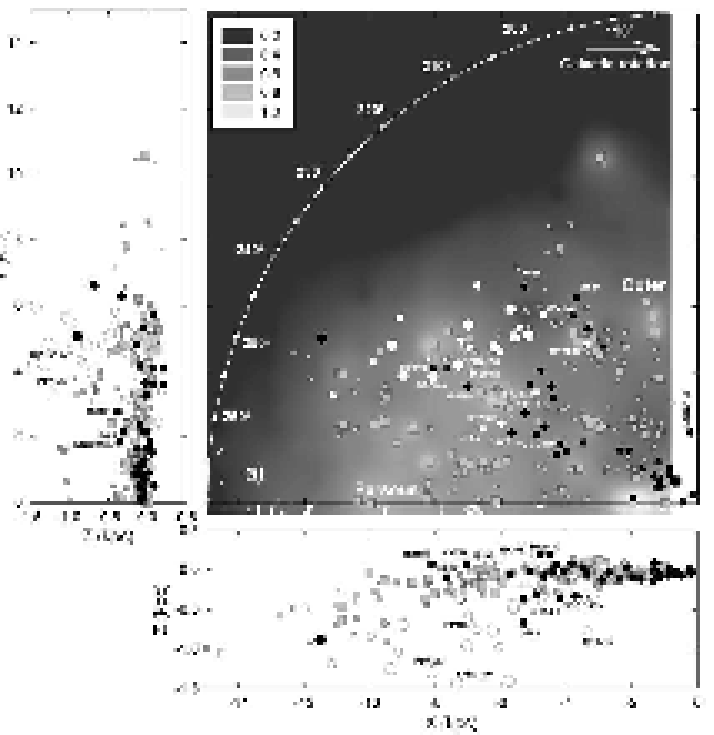}
\caption{ The CO molecular cloud mass (covering the mass ranges
$0< M_{\odot} \leq 0.5 $ -small grey squares-, $0.5 < M_{\odot}
\leq 1$ -medium sizes- and $> 1 ~M_{\odot}$ -large sizes-, in
units of $10^{5}~ M_{\odot}$) smoothing onto the X-Y plane from
May et al. (1988, 1997, 2007) data. For the smoothing process we
use the nearest neighbors procedure and a 0.5 kpc kernel. The
insert shows the mass scale of the smoothing in units of $10^{5}~
M_{\odot}$. A scale of Galactic longitudes and the Galactic
rotation are also shown. Some clusters and BPs (black and white
filled circles respectively) are indicated regarding their
positions from the formal Galactic plane at $b=0^{o}$. The lower
and left panels show the X-Z and Y-Z projections respectively of
molecular clouds, clusters and BPs. No CO smoothing was performed
in these panels. The Sun is at (0,0). Large symbols denote the new
fields observed, listed in Table 1}
\end{figure}

\clearpage

\begin{figure}
\includegraphics[scale=2.2]{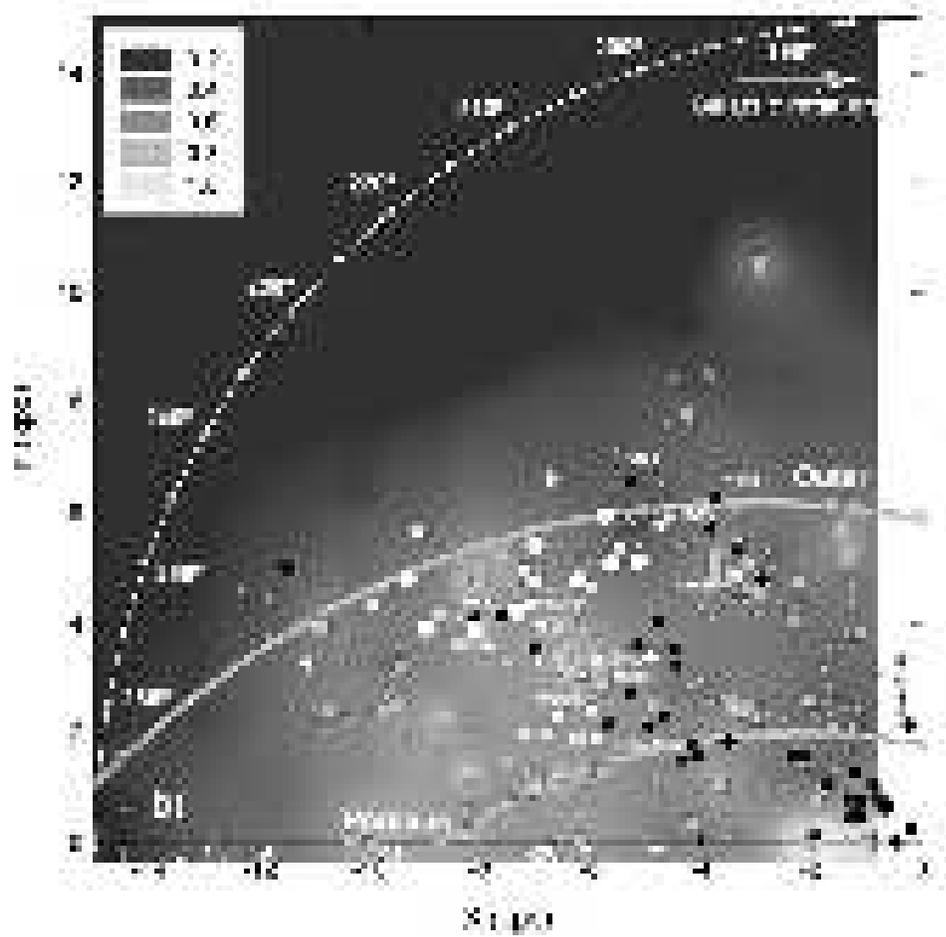}
\caption{Idem Figure 3 but including the \citet{Vallee2005} arms }
\end{figure}

\begin{figure}
\includegraphics[scale=1.2]{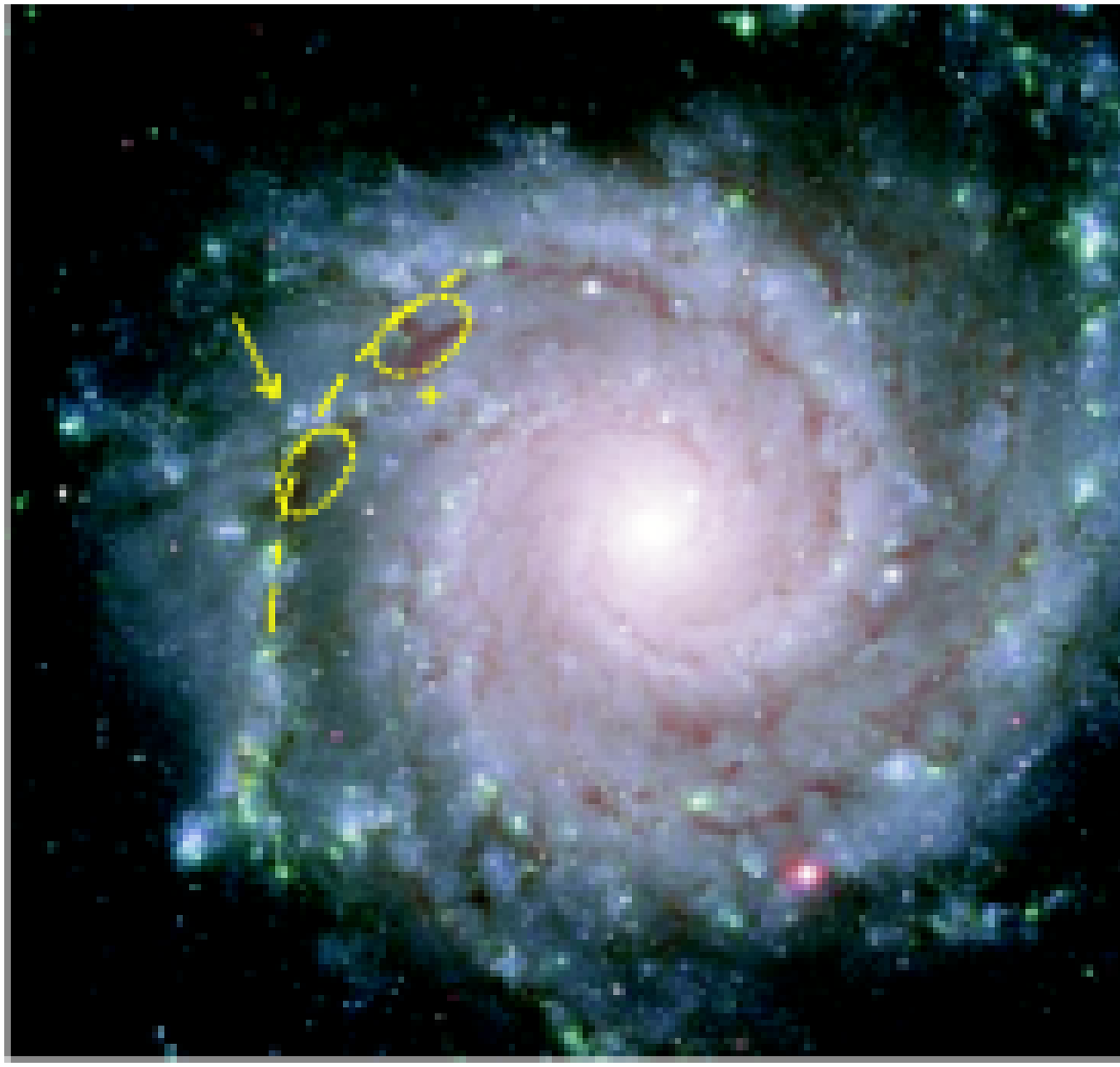}
\caption{The face-on galaxy M74 (M74 Astronomy Picture of the Day
at http://apod.nasa.gov/apod/ ap011004.html). As indicated in
\S5.3, the upper left corner sketches (long dashed yellow line) a
grand design spiral arm (like Perseus in the MW) defined by dust
tracers (dashed yellow ellipses) together with the lack of stellar
counterparts. This arm is crossed by an inter-arm spiral feature
that extends to the outer galaxy (like the Local arm in the MW).
This happens thanks to the presence of an absorption window in
between the two ellipses. We show a tentative location for the Sun
(yellow star) and the crossing region that contains an enhanced
number of young star clusters (shown by a yellow arrow)}
\end{figure}

\clearpage

\begin{figure}
\includegraphics[scale=0.6]{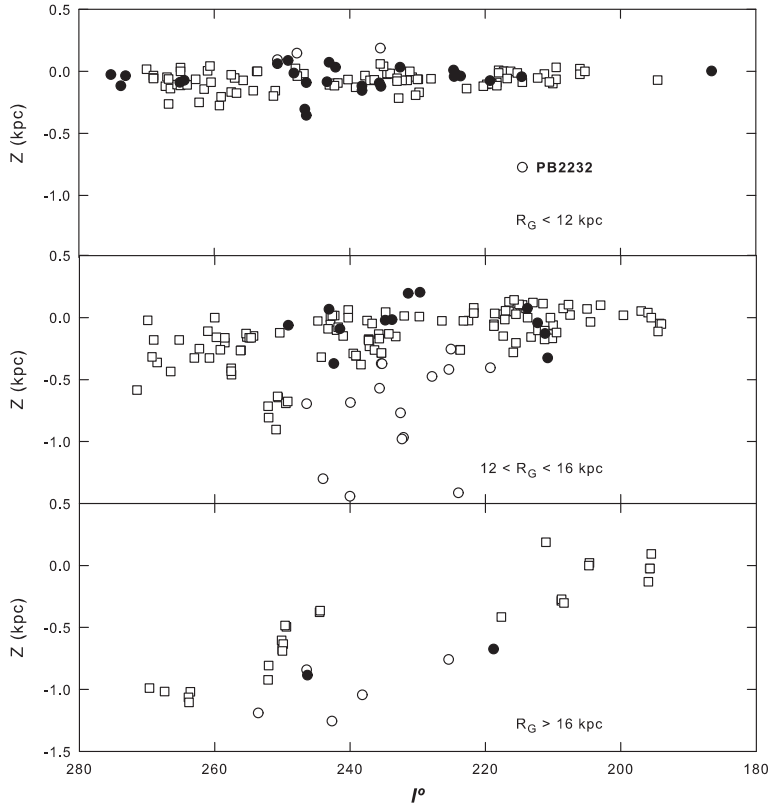}
\caption{Young open clusters (black filled circles), BPs (white
open circles) and CO clouds (open squares) in the $l-Z$ projection
distributed according to the distance from the center of the
Galaxy}
\end{figure}

\end{document}